\documentclass{moriond}

\pdfoutput=1
\usepackage{epstopdf}
\def\Journal#1#2#3#4{{#1} {\bf #2}, #3 (#4)}

\def\NPB{{\em Nucl. Phys.} B}
\def\PLB{{\em Phys. Lett.}  B}
\def\PRL{\em Phys. Rev. Lett.}
\def\PRD{{\em Phys. Rev.} D}

\def\JHEP{\em Jour. High. Ener. Phys.}

\def\be{\begin{equation}}
\def\bi{\begin{itemize}}
\def\ei{\end{itemize}}
\def\no{\nonumber}
\def\ee{\end{equation}}
\def\bea{\begin{eqnarray}}
\def\eea{\end{eqnarray}}

\newcommand{\Photo}{\includegraphics[height=35mm]{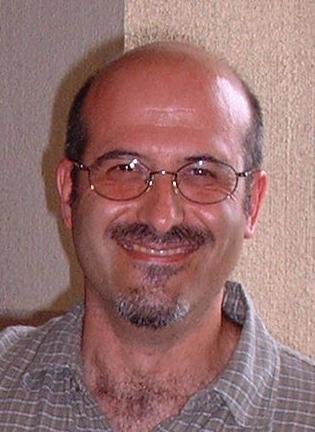}}

\begin{document}
\vspace*{4cm}
\title{RESONANCES AND UNITARITY IN COMPOSITE HIGGS MODELS}

\author{ D. ESPRIU }

\address{Institut de Ci\`encies del Cosmos, Universitat de Barcelona,\\ 
Mart\'\i ~i Franqu\`es 1, 08028 Barcelona, Spain}

\maketitle\abstracts{
The scattering of longitudinally polarized W bosons in extensions of the Standard Model 
with anomalous Higgs couplings to the gauge sector and higher order $O(p^4)$ operators is considered. 
The modified couplings should be thought as the low energy remnants of some new dynamics 
involving the electroweak symmetry breaking sector. By imposing unitarity and causality 
constraints on $W_LW_L$ scattering amplitudes we relate the possible 
values of the effective couplings to the presence of new resonances above 300 GeV. 
We investigate the properties of these new resonances and their experimental detectability.}

\section{Introduction}
We know that in the SM the Higgs boson unitarizes $W_{L} W_{L}$ scattering. Consider 
e.g. the process $W^{+}_{L} W^{-}_{L} \to Z_L Z_L$.
\begin{figure}[h!]
\centering{
\includegraphics[clip,width=0.16\textwidth]{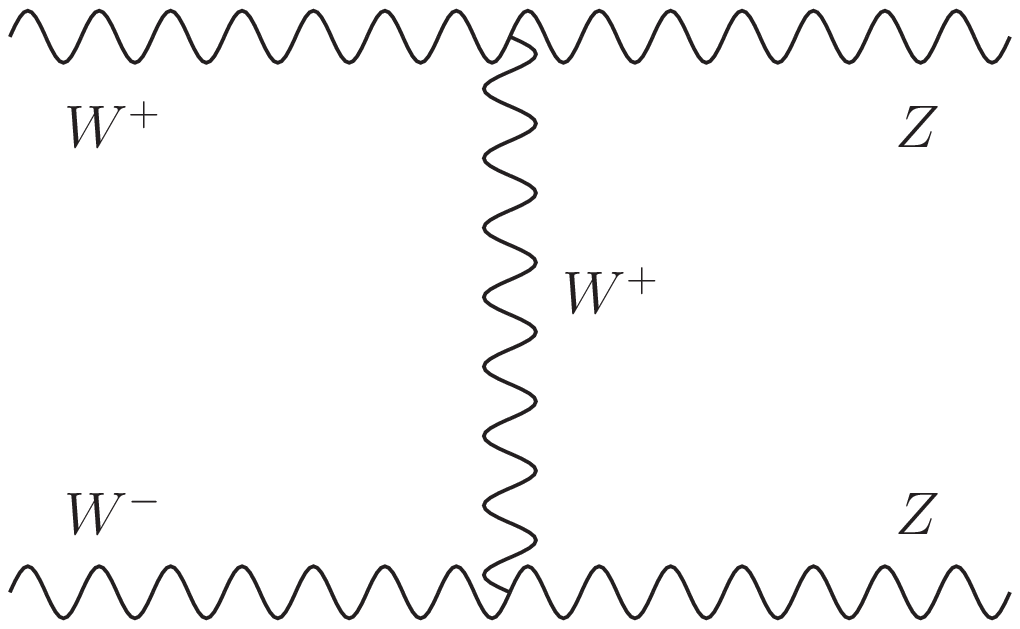} 
\includegraphics[clip,width=0.16\textwidth]{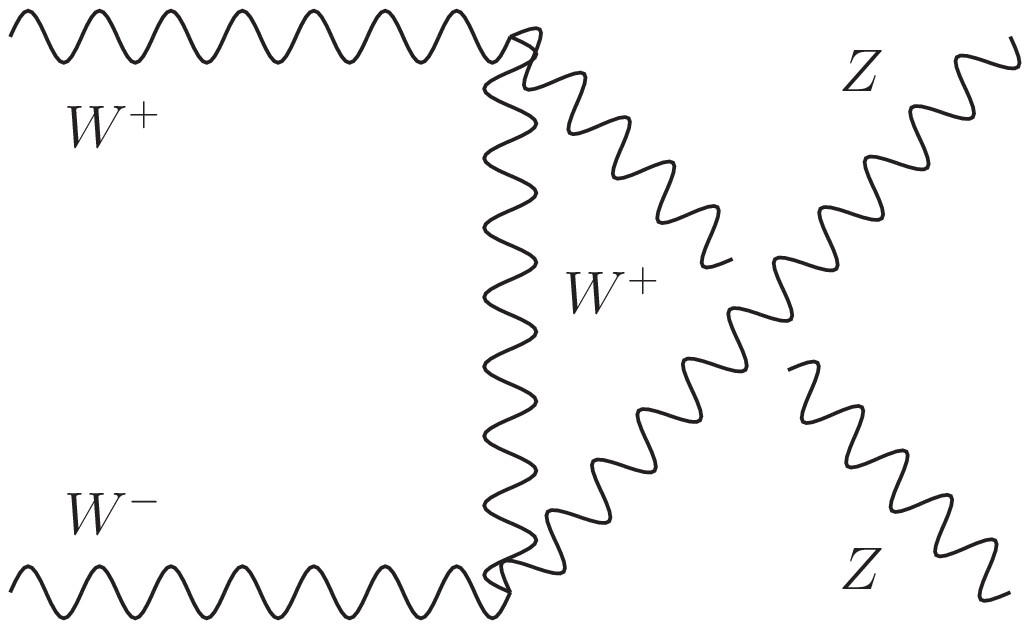}
\includegraphics[clip,width=0.16\textwidth]{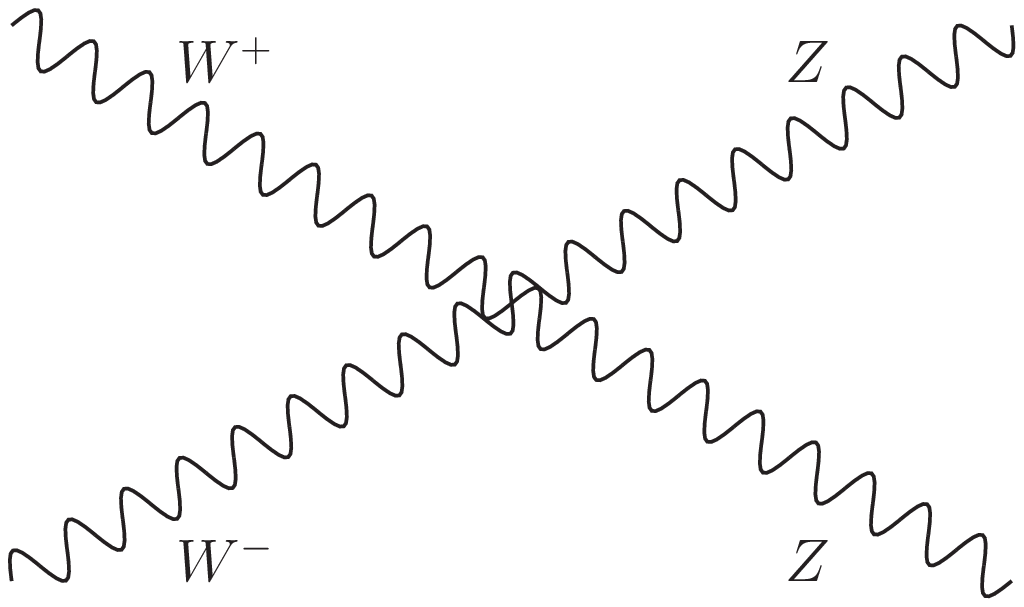} 
\includegraphics[clip,width=0.16\textwidth]{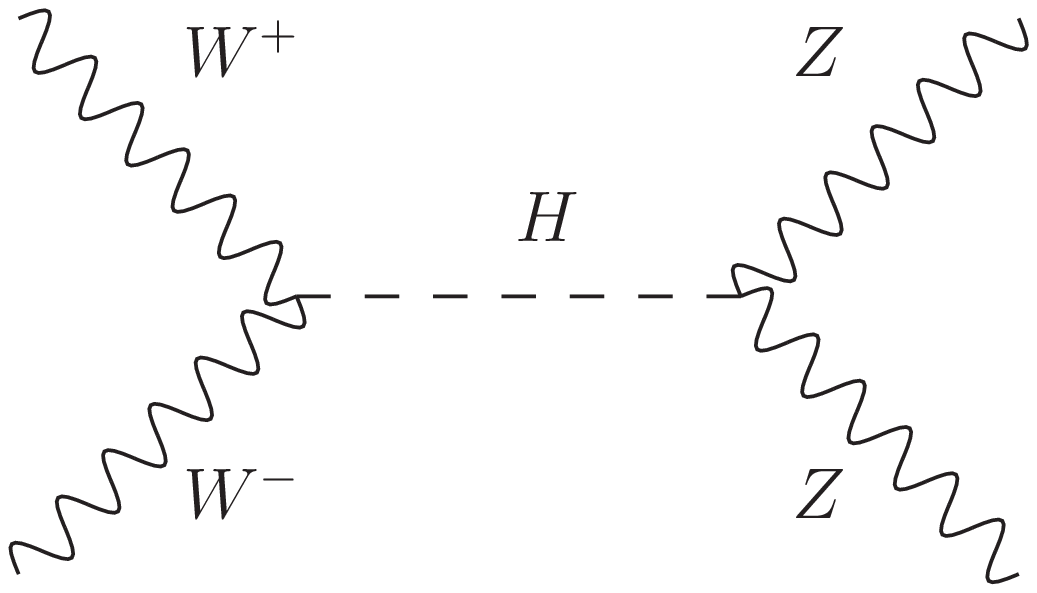}}
\end{figure}
The first 3 diagrams are fixed by gauge invariance, but we can contemplate different Higgs-gauge boson
couplings in the last one. If any of these couplings are different from the Standard Model (SM) values, the careful balance 
necessary for perturbative unitarity is lost.
For $s >> M_W^2$ the amplitude in the SM goes as
\be
\frac{s}{v^2}\frac{M_H^2}{s-M_H^2} \sim \frac{M_H^2}{v^2},
\ee
but on dimensional grounds it should go as 
\be
\frac{s}{v^2}\frac{s}{s-M_H^2}\sim \frac{s}{v^2}.
\ee
This is indeed what happens after any modification of the Higgs couplings and produces non-unitary
amplitudes. In short the SM value is precisely tuned to preserve unitarity.

Adding new effective operators typically spoils unitarity too
\be
\mathcal{L}_{SM} \to    \mathcal{L}_{SM} + \sum_i a_i \mathcal{O}_{i}.
\ee
New physics may produce either type of modifications. 
What can the requirement of unitarity in  $W_LW_L$ scattering tell us about 
possible anomalous couplings in the
electroweak (EW) sector?

\section{Parametrizing composite Higgs physics}\label{sec:parametrizing}
A light Higgs boson with mass $M_{H} \sim 125$~GeV is coupled 
to the EW bosons according to \cite{esp1}
\bea
\mathcal{L}_{\rm eff} & \supset & - \frac{1}{2} {\rm Tr} W_{\mu \, \nu} W^{\mu \, \nu} - \frac{1}{4} {\rm Tr} B_{\mu \, \nu} B^{\mu \, \nu}
+ \mathcal{L}_{\rm GF} + \mathcal{L}_{\rm FP} + \sum_{i} a_i \mathcal{O}_{i} \no \\ 
& &
+ \left[1 + 2 a \frac{h}{v} + b \frac{h}{v}^{2} \right] \frac{v^{2}}{4} {\rm Tr} D_{\mu}U^{\dagger} D^{\mu} U
 - V(h) 
\eea
\be
U =  \exp (i \; \omega \cdot \tau / v) \qquad
D_{\mu}U = \partial_{\mu}U + \frac{1}{2} i g W_{\mu}^{i} \tau^{i} U - \frac{1}{2} i g' B_{\mu}^{i} U \tau^{3} 
\ee
A non-linear realization is used. 
Setting $a=b=1$ and $a_i$=0 exactly reproduces the SM interactions.

The  $\mathcal{O}_{i}$ are a full set of $C$, $P$, and $SU(2)_{L} \times U(1)_{Y}$ gauge invariant, 
$d=4$ operators \cite{old} (of $O(p^4)$ in the chiral language) that along with the couplings $a,b$ 
parameterize the low-energy effects of an extended 
high-energy EW symmetry 
breaking sector (EWSBS) . If we assume that the EWSBS is custodially
preserving the relevant operators for $W_LW_L$ scattering are
\be
\begin{array}{ll }
\mathcal{L}_{4} = a_{4} \left({\rm Tr}\left[ V_{\mu}V_{\nu} \right] \right)^{2}, \qquad & \mathcal{L}_{5} = a_{5} \left( {\rm Tr}\left[ V_{\mu}V^{\mu} \right] \right)^{2},\qquad V_{\mu} = \left( D_{\mu} U \right) U^{\dagger}. 
\end{array}
\ee
The $a_i$ could be functions of $\frac{h}{v}$. The contribution of these $d=4$ operators to
$W_L^{(\mu)}W_L^{(\nu)}\to Z_L^{(\rho)}Z_L^{(\sigma)}$ scattering is given via the Feynman rule
\be
 i g^{4} \left[ 
a_{4} \left( g^{\mu \, \sigma} g^{\nu \, \rho} + g^{\mu \, \rho} g^{\nu \, \sigma} \right) +
2 a_{5} g^{\mu \, \nu} g^{\rho \, \sigma}
\right]
\ee

Experimentally there are by now solid indications that the Higgs particle couples to the $W, Z$ very similarly to 
the  SM rules. Let us assume for the time being that $a=b=1$ {\em exactly}. Then
\be
\mathcal{L}_{\rm eff} \simeq \mathcal{L}_{\rm SM} + 
a_{4} \left({\rm Tr}\left[ V_{\mu}V_{\nu} \right] \right)^{2} +
a_{5} \left( {\rm Tr}\left[ V_{\mu}V^{\mu} \right] \right)^{2} 
\ee
$a_{4}$ and $a_{5}$ represent anomalous 4-point couplings of the $W$ bosons 
due to an extended EWSBS that however does not manifest with $O(p^2)$ couplings 
being noticeably different to the ones in the SM. These anomalous couplings 
 will lead to violations of perturbative unitarity as they lead to amplitudes
that grow \cite{esp1,old} as $s^2$.

\section{Unitarity and resonances}\label{sec:unitarity}
Violations of unitarity are cured by the appeareance of new particles or resonances.
We can now use well-understood unitarization techniques to constrain these 
resonances and the effective couplings $\{ a_i \}$. First, let us recapitulate
  \bi
  \item The Higgs particle unitarizes amplitudes in the SM, where $a=b=1, \{a_{i}\}=0$.
\vspace{-4pt}  
   \item The theory is renormalizable without the $\{a_{i}\}$ if $a=b=1$.
\vspace{-4pt} 
  \item If present, the $\{a_{i}\}$ will then be finite non-running parameters.
  \ei
We would like to 
  \bi
  \item Determine how much room is left for the $a_i$.
\vspace{-4pt} 
  \item Find possible additional resonances required to restore unitarity.
\vspace{-4pt} 
  \item Should we have already seen any of these resonances?
\vspace{-4pt} 
  \item To what extent an extended EWSBS is excluded by current data?
  \ei
We advance some answers:
  \bi
  \item Yes, there may be new resonances with relatively light masses and narrow widths.
\vspace{-4pt} 
  \item No, we should not have seen them yet. Their signal is too weak.
\vspace{-4pt} 
  \item Looking for the resonances is an efficient (albeit indirect) way of setting constrains on a
nomalous triple and quartic gauge couplings (i.e. the $a_i $).
  \ei

Let $t_{IJ}(s)$ be a partial wave derived from the $W_LW_L\to Z_LZ_L$ amplitude. Unitarity requires
\bea
{\rm Im \;}t_{I \, J}(s) &=& \sigma(s) |t_{I \, J}(s)|^{2} \hspace{0.5cm} + \hspace{0.5cm} \sigma_{H}(s) |t_{H,I \, J}(s)|^{2} \no \\
& &  \hspace{0.5cm} \textrm{{Elastic}} \hspace{2.5cm} \textrm{{Inelastic}} \\
& &  {WW \to WW} \hspace{1.80cm}{WW \to hh} \no
\eea
\vspace{-8pt}
where $\sigma$ and $\sigma_{H}$ are phase space factors.
Given a perturbative expansion
\bea
t_{I \, J} & \approx & \hspace{0.5cm} t_{I \, J}^{(2)} \hspace{0.5cm} + \hspace{0.5cm} t_{I \, J}^{(4)} \hspace{0.5cm}+\hspace{0.5cm} \cdots \\
& &  \hspace{0.5cm} \textrm{{tree}} \hspace{1cm} {\textrm{one-loop}} \no \\ 
& &  \hspace{1.9cm} {\textrm{+~$a_{i}$~terms}} \no
\eea
\vspace{-8pt}
we can require unitarity to hold exactly by using the inverse amplitude method (IAM) to define 
\vspace{-4pt}
\be
t_{I \, J} \approx \frac{t_{I \, J}^{(2)}}{1-t_{I \, J}^{(4)}/t_{I \, J}^{(2)}} \no
\ee
for non-coupled channels.\cite{mad1} Several analyticity assumptions are implied in the above derivation. 

Unitarization of the amplitudes may result in the appearance of new heavy resonances 
associated with the high-energy theory ($
t_{00} \to  \textrm{Scalar isoscalar}  \quad
t_{11}  \to  \textrm{Vector isovector}  \quad
t_{20} \to \textrm{Scalar isotensor}$).  
We will search for poles in $t_{I \, J}(s)$ up to $4 \pi v \sim 3$~TeV (domain of applicability
of the effective theory). Physical resonances will be required to have the phase shift pass through $+\pi/2$.
This method is known to work remarkably well in strong interactions.

Is this unitarization method unique? Certainly not. Many methods exist: IAM, K-matrix approach, 
N/D expansions, Roy equations,....
While the quantitative results differ slightly, the gross picture does not change.
For a detailed discussion of the different procedures see.\cite{mad1}

\section{Calculation and results for $a=b=1$}
Most studies concerning unitarity at high energies are carried out using the Equivalence
Theorem (ET). This is understandable as calculations simplify enormously\cite{espmat}
\be
A(W^{+}_{L}W^{-}_{L} \to Z_{L}Z_{L}) \to A(\omega^{+} \omega^{-} \to \omega^{0} \omega^{0}) + O(M_W/\sqrt{s})\no
\ee
For a light Higgs one needs to include tree-level Higgs exchange as well.
Then one could  make use of the well known chiral lagrangian techniques to derive the amplitudes
and compare with experiment, including the Higgs as an explicit resonance.
However for $s$ not too large (which obviously is now an interesting region) the simplest version of the
ET may be
too crude an approximation and we shall use as much as possible exact amplitudes. 

However, a full calculation of the one-loop contribution for the $W_LW_L\to Z_LZ_L$ process, $t^{(4)}_{IJ}$,
in particular for arbitrary values of $a$ and $b$ is beyond question. Only one complete calculation exists
due to Denner and Hahn \cite{denner} for the SM case and it is available only numerically; not suitable for unitarity
analysis. We can take a shortcut.
The optical theorem implies the \textit{perturbative} relation
\bea
{\rm Im \;}t_{I \, J}^{{(4)}}(s) &=& \sigma(s) |t_{I \, J}^{{(2)}}(s)|^{2} + \sigma_{H}(s) |t_{H,I \, J}^{{(2)}}(s)|^{2} \\
\textrm{{one-loop}} & & \hspace{2cm} \textrm{{tree}} \no
\eea
For the \textit{real part}, note that
\bea
{\rm Re \,}  t_{IJ}^{(4)} &=& \textrm{$a_{i}$-dependent terms}  \quad + \quad  \textrm{real part of loop calculation} \\
& \approx & \textrm{$a_{i}$-dependent terms} \quad \textrm{(for large $s$, $a_{i}$)}\no\\ 
 \no
\eea

\vspace{-8pt}
We approximate the \textit{real part of loop contribution} with one-loop Goldstone boson amplitudes 
using the ET. The other contributions are computed exactly. See\cite{esp1} for details.

Are there resonances? To answer this questions we
must search for poles in the second Riemann sheet --- the phase shift must go 
through $+\pi/2$ at the resonance.
Are there any physically acceptable resonances? This question is answered in the positive. If one 
looks for resonances with masses below 3 TeV they  are present for 
virtually any value of $a_4$ and $a_5$, except for values very close to zero (i.e. very close to the
SM).
\begin{figure}[h!]
\centering
\includegraphics[clip,width=0.38\textwidth]{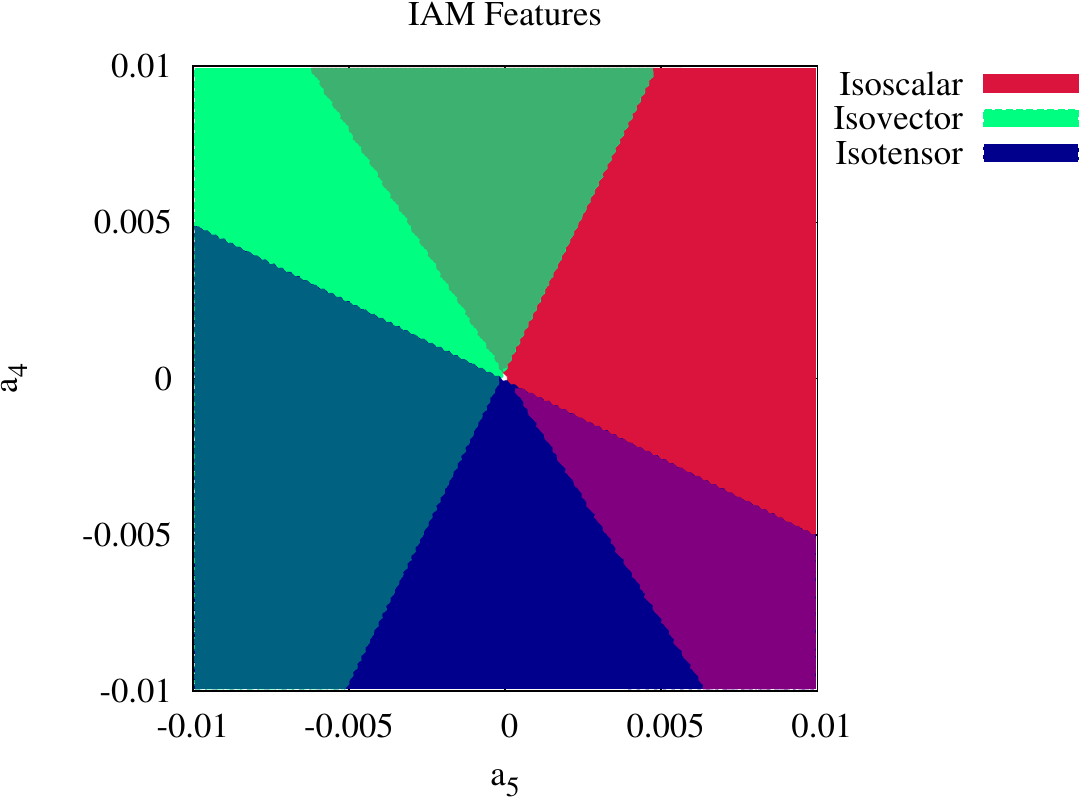} \hspace{0.25cm}
\includegraphics[clip,width=0.38\textwidth]{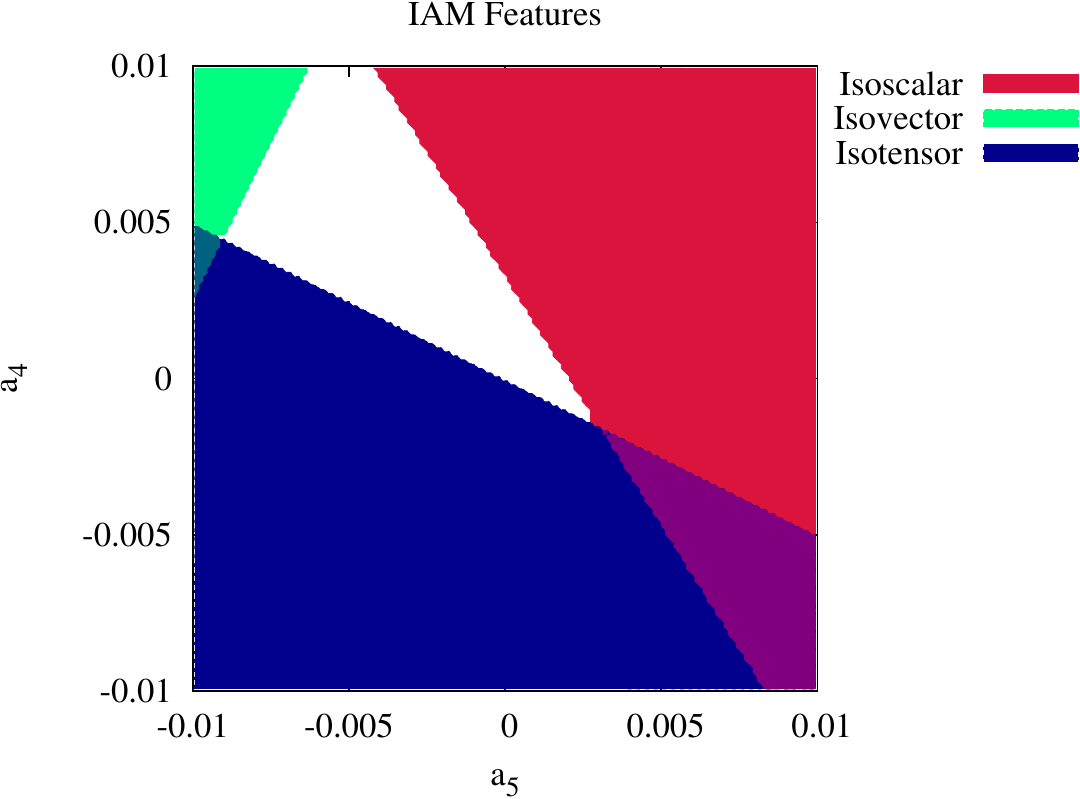}
\caption{Left: for $a=b=1$ regions in $a_4-a_5$ leading to acceptable resonances. The red (green) region
corresponds to acceptable isoscalar (isovector) resonances.
The blue-shaded area leads to acausal resonances and the corresponding values for
{ $a_4$} and { $a_5$} are unphysical --- they cannot
be realized in any effective theory with a meaningful UV completion. Only a extremely small
set of $a_4-a_5$ parameters (very close to the SM values ---zero-- nearly invisible in the figure) 
do not lead to new resonances below 3 TeV.  
Right: same but now we impose that the resonances should be found below 600 GeV. If not present, the range of values
for the anomalous couplings still acceptable (white area) is much enlarged. This could possibly 
represent the present experimental
situation according to the present analysis.}
\end{figure}

\subsection{Properties of the new resonances}
In the next figure  we show the masses that are obtained in the scalar and vector channels.
As we see, by varying the values of $a_4-a_5$ we obtain masses in the regions
$M_{S} \sim 300 - 3000$~GeV, $M_{V} \sim 550 - 2300$~GeV. This means that relatively
light masses are possible in extended EWSBS leading to appropriate values of the $d=4$
effective couplings. Observing or excluding these resonances is thus an indirect way
of measuring these effective couplings. Note that this analysis is independent of the
precise nature of this sector because only general arguments (locality, unitarity,...) have
been used.
We have similar plots for the widths but we will not present them here due to space
reasons. The resonances are generally speaking narrow:  
$\Gamma_{S} \sim 5 - 120$~GeV, $\Gamma_{V} \sim 2 - 24 $~GeV.
\begin{figure}[h!]\label{fig:masses}
\centering
\includegraphics[clip,width=0.30\textwidth]{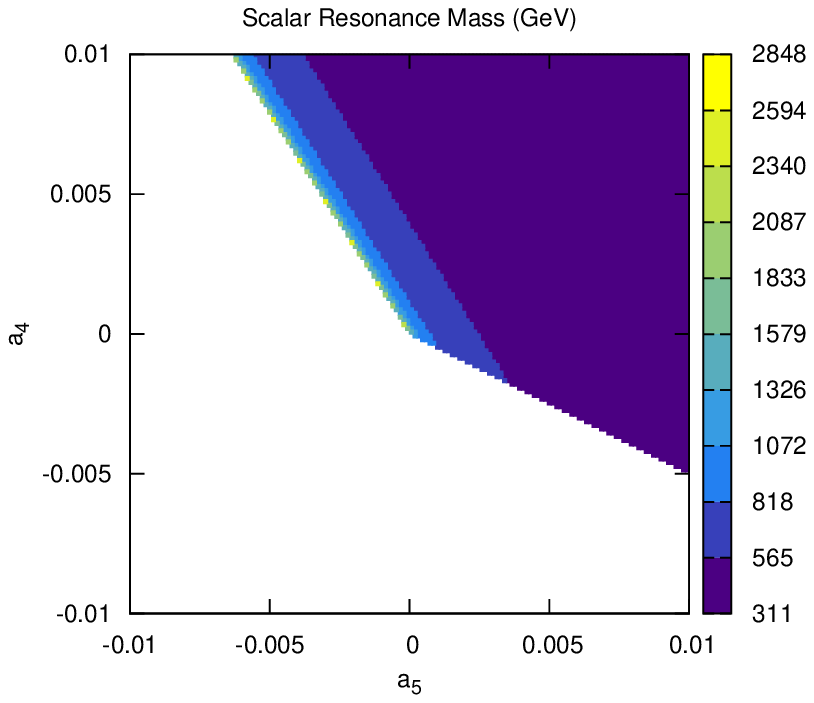} \hspace{0.5cm}
\includegraphics[clip,width=0.30\textwidth]{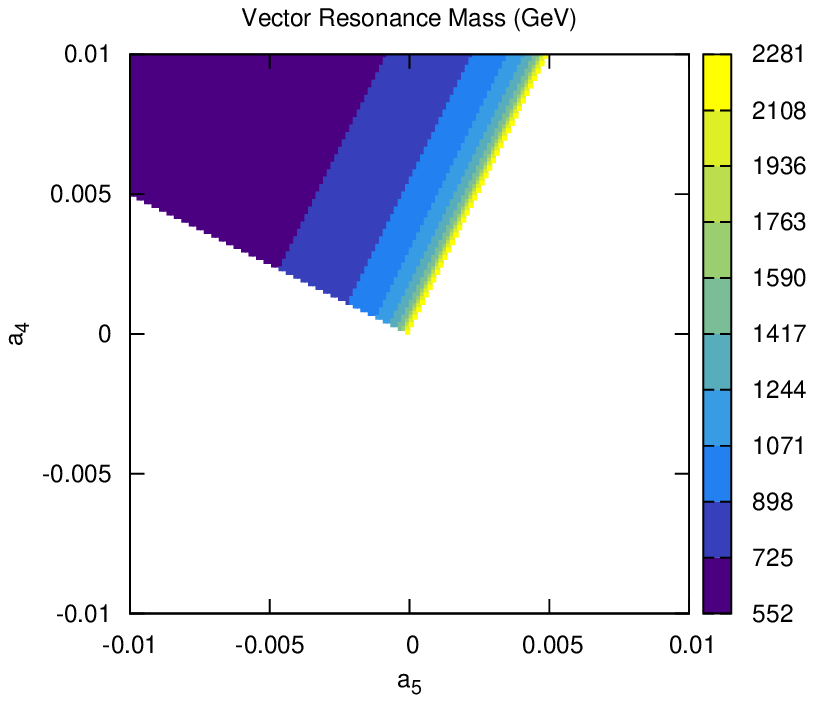} 
\end{figure}

\subsection{Visibility of the resonances}
The next question is whether these resonances are detectable. The answer is that this is impossible
with the present experimental statistics. To see this point clearly we show the signal 
of two of the resonances predicted by unitarity: one scalar and one vector. They correspond
to the values for $a_4$ and $a_5$ indicated in the figure. For these values both one scalar and
one vector resonances are present (the vector one is heavier). We compare the strength of the
signal of the scalar resonance to the one corresponding to a SM Higgs with the same mass. Resonances
could still be there, but would give a small signal. This signal is undetectable at present and will necessitate
at least 10 times more statistics. In addition this signal would only be present in the 
$WW\to WW$ or $WW\to ZZ$ channels. The large contribution that the SM Higgs represents 
leaves little room for additional resonances.
\begin{figure}[h!]
\centering{\includegraphics[clip,width=0.40\textwidth]{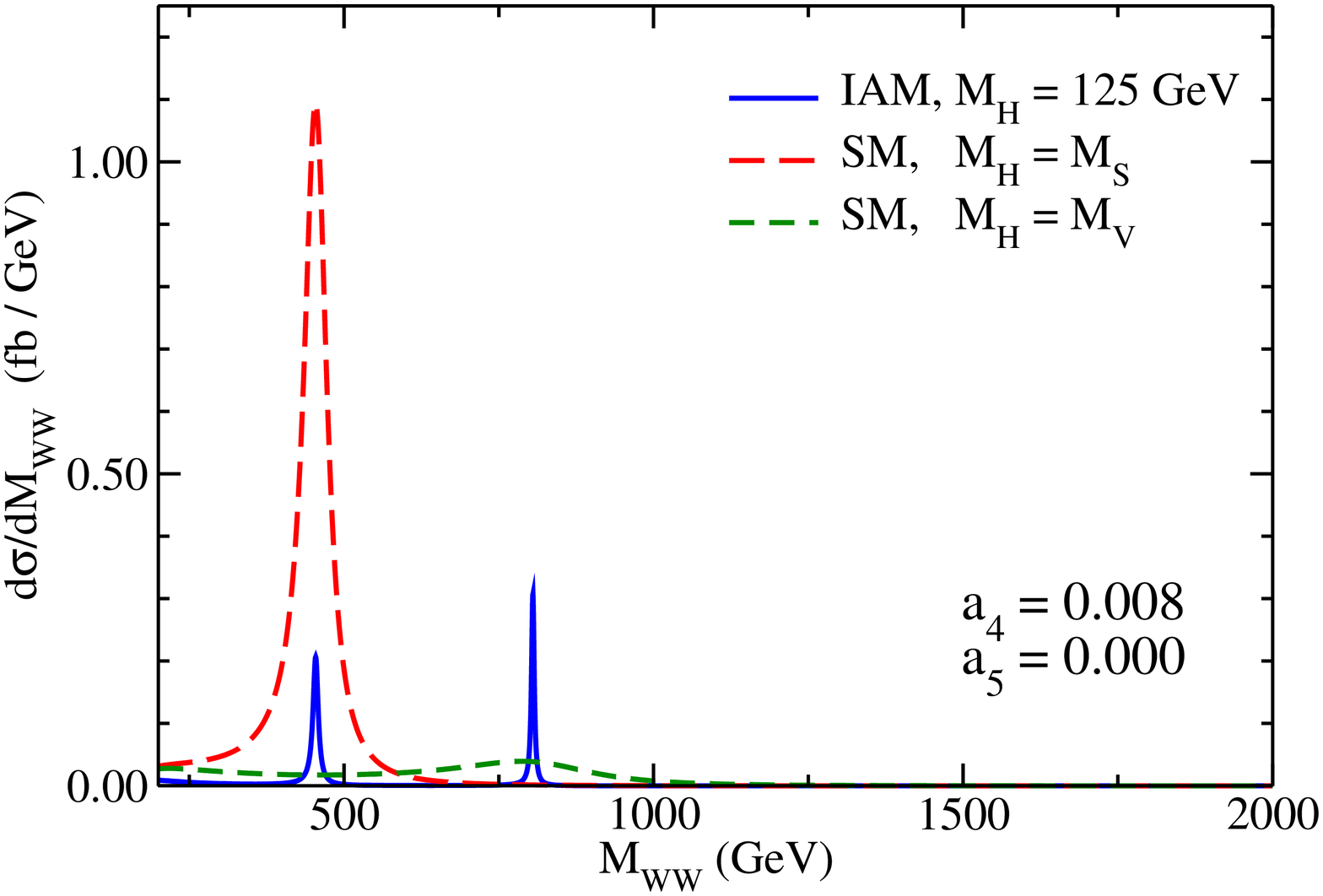}}
\end{figure}

\section{Moving away from the SM Higgs couplings}\label{subsec:moving}
What if the $hWW$ couplings are not exactly the SM ones? 
Nothing prevents us from carrying out the same programme for arbitraty values of the Higgs-to$WW$ 
couplings $a$ and $b$. 
The resulting effective theory is non-renormalizable and the $a_i$ will  be required to absorb the 
additional divergences\cite{esp2}
\be
\delta a_4= \Delta_\epsilon \frac{1}{(4\pi)^2}\frac{-1}{12} (1-a^2)^2 
\ee
\be
\delta a_5= \Delta_\epsilon \frac{1}{(4\pi)^2}\frac{-1}{24} \left[(1-a^2)^2+
\frac32((1-a^2)-(1-b))^2\right] 
\ee
We can repeat the same unitarization procedure as for $a=b=1$ and search for
resonances. The results are shown in the following figure.
\begin{figure}[h1]
\centering{
\includegraphics[clip,width=0.38\textwidth]{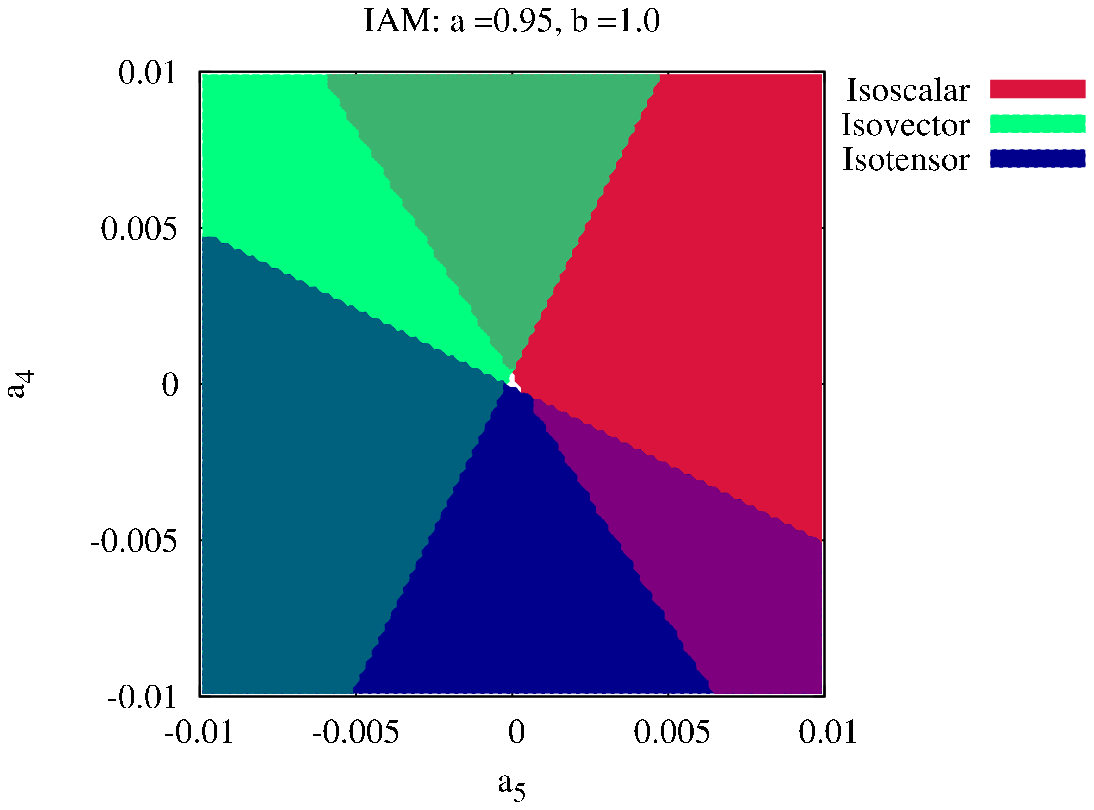} \hspace{0.15cm}
\includegraphics[clip,width=0.38\textwidth]{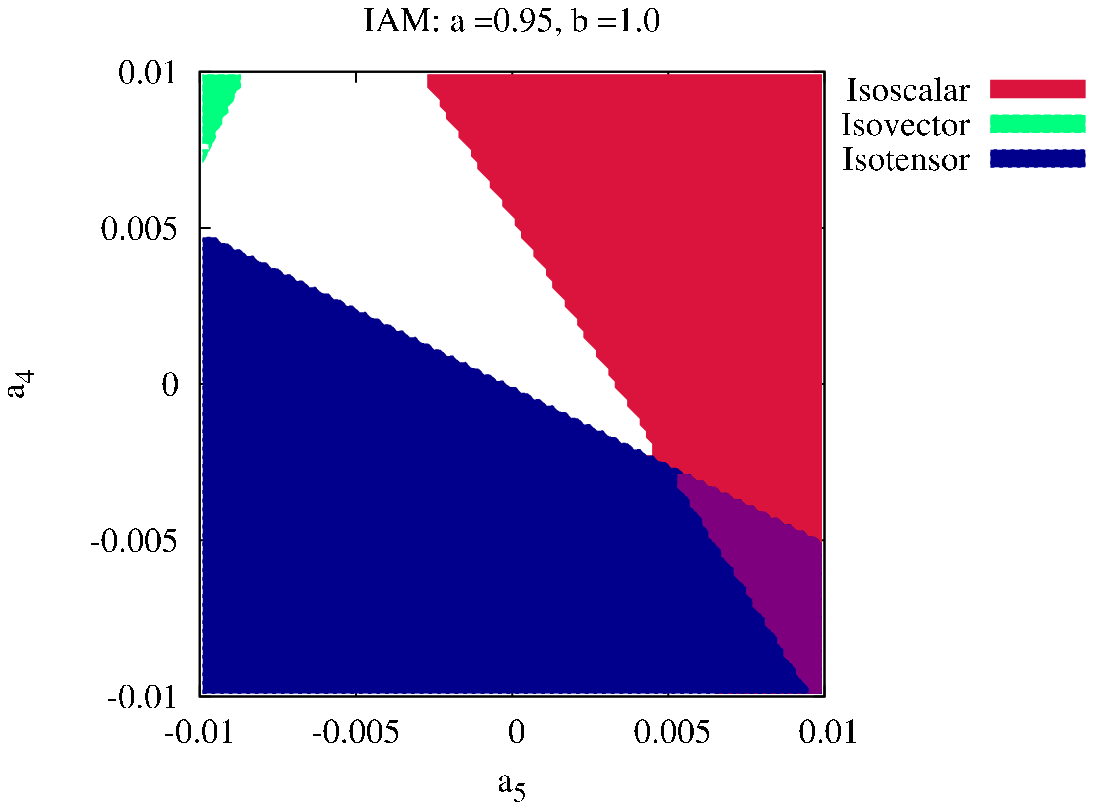}}
\caption{Exclusion zones and bounds on $a_i$ for $a=0.95$ ($b=a^2$). Left: in color the excluded region 
in $a_4-a_5$ parameter space
if no resonance is found $< 3$ TeV. Right: in color the excluded regions if no resonance 
exists below $M_{R}=600$~GeV. Note that the allowed white areas are slightly larger than in the case $a=b=1$}
\end{figure}
The characteristics of the resonances tend smoothly to the $a=1$ case ($hWW$ coupling as in
the SM). Resonances tend to be slightly heavier and broader than for {$a=1$}.
The parameter {$b$} is only marginally visible in the widths (not shown).
There are constraints on vector masses from $S,T,U$ parameter constraints in some models.\cite{pich}

As in the {$a=1$} case the signal is always much lower\cite{esp1} than the one for a Higgs of the same mass.
For {$a=1$} typically $\sigma_{\textrm{resonance}}/\sigma_{\textrm{Higgs}} < 0.1 $, 
now  $\sigma_{\textrm{resonance}}/\sigma_{\textrm{Higgs}} \simeq 0.2 $.

To sumarize, the situation for $a<1$ is not radically different from $a=1$. 
Resonances (particularly in the vector channel) are slightly more difficult to appear.
They tend to be slightly heavier and broader and
they give a slightly larger experimental signal.

This situation changes drastically for $a>1$. `Something' happens when {$a>1$}. Most of the resonances 
disappear and in fact most of parameter space is excluded on causality and unitarity grounds. We have no space left
to explain the reasons of this radical change of behaviour here and recommend the interested reader to
examine our references.\cite{falk,esp1} From a technical point of view, this drastic modification is associated to the
change of sign of $t^{(2)}$ when $a> 1$.

Let us summarize our main points. Unitarity is a powerful constraint on scattering amplitudes. The validity 
is well tested in other physical situations.
Even in the presence of a light Higgs, unitarization can help constrain anomalous couplings by 
helping predict heavier resonances.
An extended EWSBS would typically have such resonances even in the presence of a 125 GeV Higgs. 
However the properties of the resonances are radically different from the `standard lore'.
Limited by statistics, existing LHC searches do not yet probe the IAM resonances.

\section*{Acknowledgements}
This work is supported by grants FPA2013-46570, 2014-SGR-104 and Consolider grant CSD2007-00042 (CPAN). It 
is a real pleasure to thank my collaborators B. Yencho and F. Mescia.

\end{document}